\documentclass[pdflatex,sn-mathphys-num]{sn-jnl}

\usepackage{graphicx}%
\usepackage{multirow}%
\usepackage{amsmath,amssymb,amsfonts}%
\usepackage{amsthm}%
\usepackage{mathrsfs}%
\usepackage[title]{appendix}%
\usepackage{xcolor}%
\usepackage{textcomp}%
\usepackage{manyfoot}%
\usepackage{booktabs}%
\usepackage{algorithm}%
\usepackage{algorithmicx}%
\usepackage{algpseudocode}%
\usepackage{listings}%

\theoremstyle{thmstyleone}%
\theoremstyle{thmstyletwo}%

\theoremstyle{thmstylethree}%

\begin{document}

\title[Article Title]{Optimizing QUBO generation parameters for NP problems and their impact on D-Wave convergence}


\author*[1]{\fnm{Toru} \sur{Fujii}}\email{Toru.Fujii@nikon.com}

\author[1]{\fnm{Koshi} \sur{Komuro}}\email{Koshi.Komuro@nikon.com}

\author[1]{\fnm{Kaito} \sur{Tomari}}\email{Kaito.Tomari@nikon.com}

\affil[1]{\orgdiv{Fundamental Technology Development Department}, \orgname{Nikon Corporation}, \orgaddress{\street{1-5-20 Nishioi}, \city{Shinagawa-ku}, \postcode{140-8601}, \state{Tokyo}, \country{Japan}}}




\abstract{NP problems are closely related to practical optimization problems but often face exponential increases in computation time as problem sizes grow, prompting numerous attempts to accelerate calculations. Quantum annealing has been explored as a promising approach to solve NP problems faster than classical computers. However, it requires expressing cost functions and constraints as an energy function in QUBO format. Parameter settings for QUBO coefficients are often empirical, especially in scheduling problems, where large values may be required, demanding experience and intuition. In this study, we analyzed QUBO generation formulas for three problems classified as coloring problems in Lucas's paper: the graph coloring problem, the clique vertex cover problem, and the integer-length job scheduling problem. We identified the necessity of independent parameters for complex problems. By analyzing QUBO states and eigenvalues from modified formulas, we derived relationships between formula characteristics and optimal QUBO parameter values, along with their calculation methods. Using the quantum annealing machine D-Wave, we validated the derived parameters. Additionally, we visualized the impact of parameter changes on states and eigenvalues using small spin problems. We also demonstrated the existence of independent Ising coefficients that enhance convergence to correct states, depending on optimal parameter changes for ground-state and non-ground-state problems.}




\maketitle

\section{Introduction}\label{sec1}

``Cook's Theorem'' (``The Complexity of Theorem-Proving Procedures''), published by Stephen Cook in 1971 \cite{Cook71}, introduced the concept of NP-completeness and represented a groundbreaking achievement in the study of NP problems. Richard M. Karp's ``Reducibility Among Combinatorial Problems'' (1972) \cite{Karp72} is also a historic paper in computational complexity theory, introducing 21 NP-complete problems, which sparked the spread of the concept of NP-completeness. Since then, NP problems have been actively studied \cite{Garey79}, and are known to be closely related to various practical optimization problems \cite{Yarkoni22} , such as scheduling problems \cite{Pinedo16}. In recent years, NP problems have been used not only in the field of classical calculations, but also in quantum chemical calculations \cite{Verteletskyi20}, and NP problems such as graph coloring and minimum clique problems are used to improve the efficiency of calculations. The applied importance of NP problems is increasing.

Representative NP problems include the Traveling Salesman Problem (TSP), the Graph Coloring Problem (GCP), and the Satisfiability Problem (SAT). Problems with practical applications often include the knapsack problem, scheduling problems such as job shop scheduling, and facility location problems \cite{Garey79}.  A well-known characteristic of NP problems is that the calculation time generally grows exponentially with the problem scale. As a result, there are countless studies on how to speed up calculations.
With the emergence of D-Wave \cite{Johnson11}, a quantum annealing (QA) machine that is expected to solve optimization problems faster than classical computers, adiabatic quantum computing \cite{Farhi00} and quantum annealing have begun to be applied to NP problems. Some NP problems can be expressed in QUBO \cite{Barahona82} \cite{Lucas14}, with cost functions and constraints as energy functions. 

With the aim of broadly applying QA to NP problems, Lucas provided Ising formulations for NP-complete and NP-hard problems \cite{Lucas14}, including Karp's 21 NP-complete problems \cite{Karp72}. As attempts are being made to apply quantum annealing, it is being actively applied not only to quantum but also to various Ising machines \cite{Mohseni22} \cite{Zhang24}. There are also reports of partially good results being obtained.

A polynomial time solution for one NP-complete problem implies that there are polynomial time solutions for all NP-complete problems, but in practice, detailed ingenuity is needed to consistently get the speed up, depending on the problem. One of the things that is known is that the solution space increases when expressing a problem as a QUBO, so it is necessary to make QUBO more efficient \cite{Glover18} \cite{Boros02} , to embed the QUBO into D-Wave graph efficiently \cite{Date19} and research is being conducted to achieve this.

We use the coloring problem in Lucas Section 6 as an example, and calculate the optimal parameters by reviewing each term in the QUBO generation formula, and verify the convergence characteristics of the QUBO generated using the optimal parameters with D-Wave. First, we calculate the numerical values that are thought to be related to the annealing convergence, and determine the dynamic range as the ratio of the eigenvalue difference near the ground state to the maximum and minimum eigenvalue differences.  We then determine the coefficient parameters to reduce the dynamic range. We numerically show cases where the handling of Lucas's coefficient parameters is sufficient and cases where improvements are possible. We generate a QUBO near the predicted optimal parameters, and use D-Wave to evaluate the probability of convergence to the ground state, a state in which the problem itself can be considered to be solved with the correct answer. We also visualize and understand the change in convergence characteristics by investigating the effect of parameter changes on the state and eigenvalues of the generated QUBO. Finally, we show that there are characteristic Ising coefficients that change the convergence characteristics.

\section{NP problems for which the correct answer is obvious}\label{sec2}

We select NP problems from Lucas's paper. Lucas broadly classifies them into eight types, ranging from two-section partitioning problems to nine-section graph isomorphism problems. Among them, we use the 6-section coloring problem type, which includes problems that are influenced by the size of the elements handled in the problem.  These serve as examples to show the basic policy for parameter determination and the results of running the problem on a D-Wave QPU. Lucas's constraints for the three problems in the six sections are shown below.

The graph coloring problem, GCP in Section 6.1 of the Lucas paper is expressed by Eq. (1) or Eq. (51) in the Lucas paper. In the case of planar graphs, it is widely known as the 4-coloring problem of a plane map, and it is known that it is already NP-complete when the color partition is three colors:
\begin{equation}
H = A \sum_{v} \left( 1 - \sum_{i=1}^{n} x_{v, i} \right)^2 + A \sum_{(uv) \in E} \sum_{i = 1}^{n} x_{u, i} x_{v, i}.
\label{eq1}
\end{equation}
In the formula, $A$ represents a parameter used to adjust the weight of each term, while $u$ and $v$ denote nodes, $i$ indicates a color, $n$ represents the total number of colors, and $E$ corresponds to an edge set.

The clique vertex cover problem (CVCP) in Section 6.2 is also expressed by equation (2) similar to the graph coloring problem, and the energy equation in the Lucas paper (52). Both are simple problems, but they have recently been used to make quantum chemical calculations more efficient:
\begin{equation}
H = A \sum_{v} \left( 1 - \sum_{i=1}^{n} x_{v, i} \right)^2 + B \sum_{i = 1}^{n} \left[ \dfrac{1}{2} \left( -1 + \sum_{v} x_{v, i} \right) \sum_{v} x_{v, i} - \sum_{(uv) \in E} x_{u, i} x_{v, i} \right].
\label{eq2}
\end{equation}
In the formula,  $A$ and $B$ represent parameters used to adjust the weight of each term, while $u$ and $v$ denote nodes, $i$ indicates a color, $n$ represents the total number of colors, and $E$ corresponds to an edge set.
Job sequencing with integer length in Section 6.3 is also called the identical parallel machine scheduling problem and is one of the simplest scheduling problems. It is expressed as the sum of $H_{A}$ and $H_{B}$:
\begin{equation}
H_{A} = A \sum_{i = 1}^{N} \left( 1 - \sum_{\alpha} x_{i, \alpha} \right)^2 + A \sum_{\alpha = 1}^{m} \left[ \sum_{n = 1}^{\mathcal{M}} n y_{n, \alpha} + \sum_{i} L_{i} \left( x_{i, \alpha} - x_{i, 1} \right) \right] ^2
\label{eq3}
\end{equation}
and
\begin{equation}
H_{B} = B \sum_{i} L_{i}  x_{i, 1} ,
\label{eq4}
\end{equation}

In the formula, $H_{B}$ represents the longest process time among all the machines, which is minimized under the constraint equation $H_{A}$. The variables are defined as follows: $A$ and $B$ are parameters used to adjust the weight of each term, $i$ denotes a job, $N$ is the total number of jobs, $\alpha$ represents a machine, $m$ is the total number of machines, $n$ refers to the process time difference between the machines with the longest process times and others, $\mathcal{M}$ is the maximum difference time set, and $L$ indicates the process time of each job.

Although it is the simplest of the scheduling problems, compared to GCP and CVCP, where the edge weights are equal and the job lengths are equal, the process time of PMSP jobs has a natural number of degrees of freedom. As will be discussed later, the coefficient parameters of each term in the QUBO constraint equation are highly dependent on the job process time. However, we will show that it is possible to determine the parameters from the QUBO energy equation. In addition, when the parameters are appropriately selected, it is the first term that determines the minimum makespan, and the third term of the constraint equation exists in a state that represents the minimum makespan even if it is not the ground state spin configuration. In other words, we will show later that even a set of spins with incorrect constraint conditions that are not in the ground state contains a minimum makespan, that is, a state that can be used as a correct answer in practice, which we call a practical correct answer state. It will also be shown that by appropriately selecting the parameters, the convergence probability of the practical correct answer increases in the same way as the ground state.

\begin{figure}[b]
\centering
\includegraphics[width=0.9\textwidth]{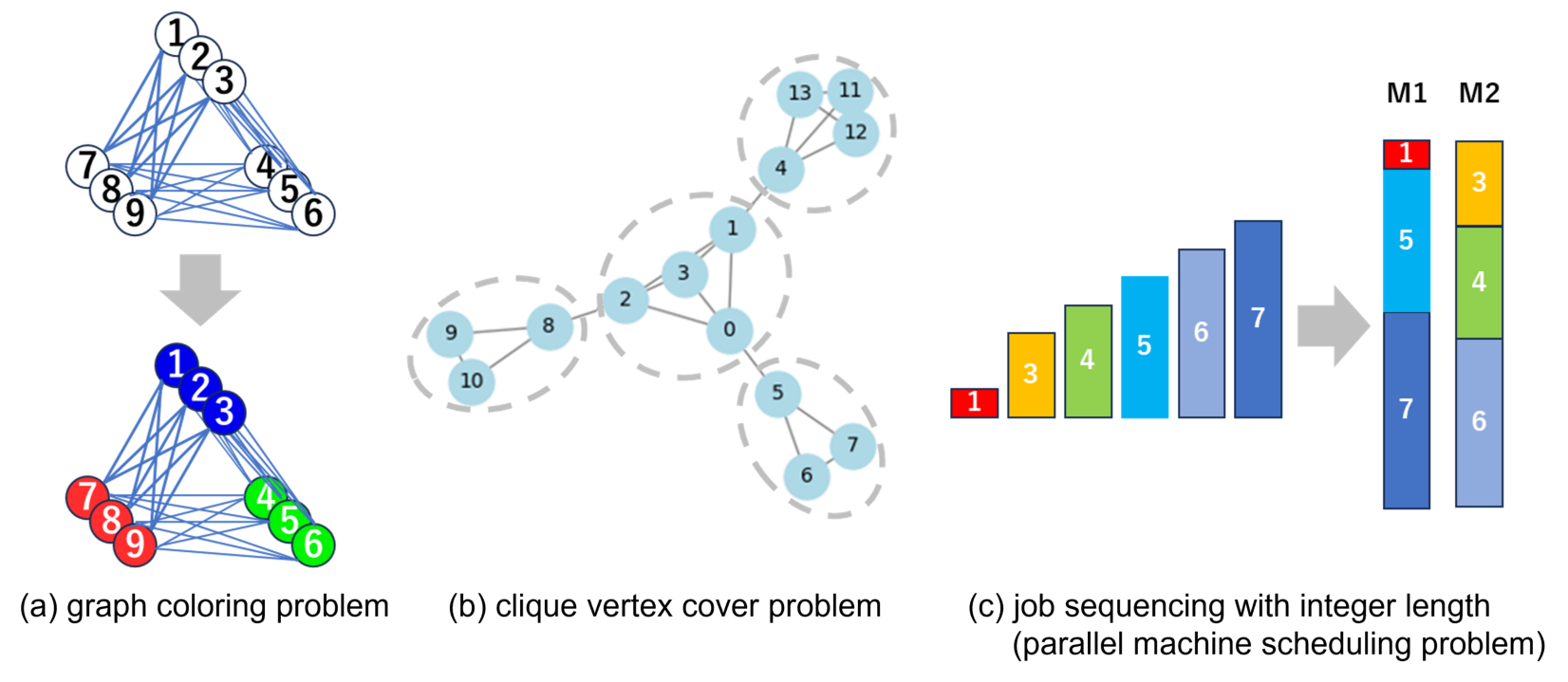}
\caption{(a) top: graph coloring problem using complete three-partite graph, (a) bottom: minimum coloring example, (b) trivial clique vertex cover problem, (c) job sequencing with integer length, 2-machine 6-job parallel machine scheduling problem (PMSP) as a scheduling problem.}
\label{fig1}
\end{figure}

Examples of these three types of problems with obvious answers are shown in Fig. \ref{fig1}.
Since the graph coloring problem has a restriction that nodes connected by edges must not have the same color, a coloring like the minimum coloring example below will have the lowest QUBO energy ($=0$). In order to make the correct answer self-evident, a 3D version of a complete Department graph like the one shown in Fig. \ref{fig1} is used for this problem analysis and D-Wave evaluation. The center is a clique vertex cover problem, and dividing it into four cliques surrounded by gray dotted lines will have the lowest QUBO energy ($=0$). Since the clique vertex cover problem can only be solved with about 100 spins on D-Wave's QPU, the limit is a 5-color problem with 20 nodes, so a self-evident problem is created by preparing multiple cliques and connecting the cliques with only one edge as shown in the figure. On the right in Figure 1 is job sequencing with integer length, which is mathematically the same as the parallel machine scheduling problem (PMSP). The problem is the example on the left of allocating six jobs to two machines. The correct answer is the machine assignment of the job that has the shortest process time on the last machine to finish. In this case, the problem is set so that the two machines finish simultaneously, since it is trivial to assume that the problem is the shortest. The procedure for creating trivial problems for PMSP will be shown separately.

When evaluating these problems with QA problems with self-evident solutions are not only readily verifiable, but they are also useful to evaluate the nature of computational speed ups as a function of parameter settings. It is particularly convenient for the correct answer to be self-evident when creating and evaluating large-scale problems. As already explained, it is relatively easy to create self-evident problems for the graph coloring problem and the clique vertex cover problem. Next, we will look at the properties of the problems that can be read from the QUBO generation formula and the constraint formulas therein.

For the graph coloring problem, we use a problem that can be considered a Complete three-partite Graph as shown in  Fig. \ref{fig1}(a). If the chromatic number is always 3 and a QUBO is generated, there will be color swaps between 3-color groups, but coloring is trivial. For a problem where the correct answer is three colors, the minimum unit is $N=3$, a triangle with three nodes. The figure shows the case of $N=9$. By deciding to connect the nodes of the triangle in a clockwise direction from the node itself, edges extend from each of the $N/3$ nodes to the $N/3$ node, so the total number of edges is $N/3 \times N/3 \times 3=N^2/3$. In general, for k colors, $N-N/k=N(1-1/k)$ edges will come out of $N/k$ nodes, and by dividing the overlaps by 2, the total number of edges is $N/k \times N(1-1/k) \times k/2 = N^2 \times (1-1/k)/2$. When $k=2,3$, $N/2\times N/2=N^2/4$ and $N/3\times N\times 2/3\times 3/2=N^2/3$, respectively.

The first term in Eq. (\ref{eq1}) is the penalty term for assigning one color to one node, and the second term is the penalty term when nodes connected by an edge are painted the same color, and the correct answer is when both terms are $0$. Lucas uses a common weight parameter $A$ for the two terms.

The clique vertex cover problem is trivially solved by preparing multiple cliques and generating them by connecting two of them with a single edge, as shown in Fig. \ref{fig1}(b). Let's take a quick look at the role of Eq. (\ref{eq2}). The first term is a penalty term that limits the number of colors assigned to one node, the first term in the second term is the number of edges when the node is a clique when specified, and the second term is the number of edges between the extracted nodes, which is zero only when the extracted nodes form a clique, and a positive integer when they do not. In other words, just like the graph coloring problem, the correct answer is when both terms are $0$. Here, Lucas uses separate weight parameters for the two terms.

When creating a parallel machine scheduling problem, a little more care is required than with the previous two problems. As with the previous two problems, the answer is self-evident, but in addition, a condition is imposed that makes it impossible to solve using the greedy method, which is a simple one-time sorting. If the answer is not self-evident, it will take exponentially longer to obtain the answer required to confirm ground-state convergence when the problem becomes large-scale. Also, when problems that can be solved using the greedy method are compared with the classical calculation time, the classical method, which can be solved with one sorting, will always win.

\begin{figure}[t]
\centering
\includegraphics[width=0.5\textwidth]{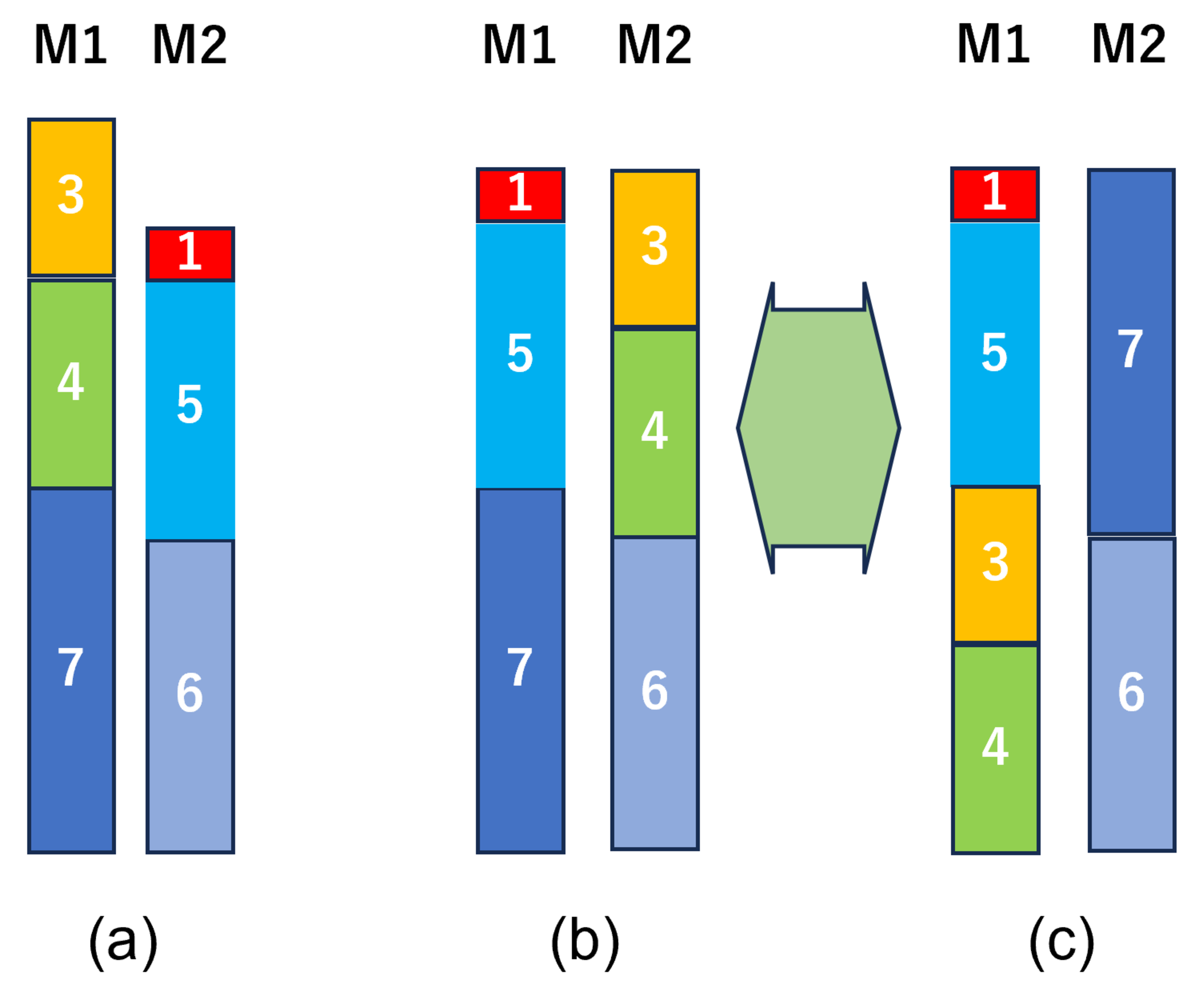}
\caption{(a) The result of solving the problem in Fig. \ref{fig1}(c) using the greedy algorithm. (b) The correct solution with the smallest makespan, showing the steps to create an even makespan problem. (c) Another correct solution with the smallest makespan. The center and left also show the correct solution when swapping all jobs between $M_{1}$ and $M_{2}$.}
\label{fig2}
\end{figure}

To make a problem unsolvable by the greedy method, you can create it using the following procedure, as shown in Fig. \ref{fig2}(a). Here, we take a problem with 2 machines and 6 jobs as an example, but in the case of 2 machines, we can consider these 6 jobs as the last 6 jobs from the smallest and arrange the even jobs up to that point to give all machines equal process times, thereby increasing the number of jobs. However, if the number of jobs is increased, the probability that the number of combinations with the same total process time per machine will increase further. The ground state will degenerate further and the number of correct solutions will increase, making it probabilistically slightly easier to search for the ground state than for a problem without degeneracy. Figs. \ref{fig2}(b) and (c) show an example of degenerate.

The jobs assigned to the two machines on the left in Fig. \ref{fig2}(a) were assigned using a greedy method, and it shows that the problem described in Fig. \ref{fig1}(c) cannot be solved correctly using a greedy method. To explain the procedure step by step, first, a job with a process time of 7 is loaded on the bottom left machine $M_{1}$, and a job with a process time of 6 is loaded on $M_{2}$ in descending order. The rule for loading from the second time onwards is to load the job with the longest process time among the jobs not loaded on the machine with the shortest total job time up to that point, so 5 is loaded on $M_{2}$ with a total job time of 6, and 4 is loaded on $M_{1}$ with a total job time of 7. Since the two machines are even at this point, it doesn't matter which machine you load the remaining jobs on in the end, but the makespan will be 14, resulting in the machine with the longest process time. Figure. \ref{fig2}(b) shows the procedure for creating the problem, and when loading for the second time, the machine with the longest total job time loads the job with the longest process time among the jobs not loaded, and when loading the last third job, the process times of 3 and 1 are adjusted so that the total job times of the two machines are equal.

Next, consider the degeneracy of the problem. As shown in Fig. \ref{fig2}(c), the job with process time 7 in $M_{1}$ can be combined and exchanged with jobs with process times 3 and 4 in $M_{2}$, and since there is not a single correct answer, the number of ground states is greater than in a non-degenerate problem. Even in the case of a problem with equal total process times, the correct job combination is always $m!$, the total number of
machines, times greater because swapping between m machines is also a correct answer, so the minimum degeneracy is $m!$.

The parallel machine scheduling problem is more complicated than the previous two problems. The edges of the two problems were unweighted and uniform up until now, but the process time of the jobs, which is an element of the parallel machine scheduling problem, is not uniform. If they are uniform, the problem becomes trivial, so there is no point in solving it. This makes the contents of the constraint equations more complex. Furthermore, the correct answer to the constraint equation, which was previously zero when the correct answer was given, and becomes a positive integer indicating the minimum makespan value when $H_B$ in Eq. (\ref{eq4}) is minimum, is now problem-dependent. The first term of $H_A$ in Eq. (\ref{eq3}) is a penalty term when a job is not assigned to a single machine. The second term is a penalty term that selects a specific machine among multiple machines and requires that the process time of that machine be equal to or longer than that of any other machine. The total process time for the last machine to finish, which is obtained by minimizing the total process time $H_B$ for the machine with the longest process time, assuming that these penalty terms of $H_A$ work well, is called the makespan, and this is the minimum. Lucas's paper also describes the conditions for the ground state to match the minimum makespan, but here we will discuss in more detail the conditions and values for obtaining the optimal values of the coefficient parameters.

\section{Basic principles for coefficient parameter setting and judgment methods}\label{sec3}
What became clear through solving the problem with D-Wave was that, apart from the quantum annealing, the system has characteristics that can be explained by the characteristics of classical analog electronic device. The dynamic range, which is the difference in eigenvalues near the ground state divided by the difference between the maximum and minimum eigenvalues, which is also a characteristic of classical annealing, has a large influence on the probability of convergence to the ground state.

First, from each term of the QUBO generation formula corresponding to the problem type, we identify the eigenvalue difference near the ground state, the state where the total energy is at or near the maximum eigenvalue, and the state where each term is at its maximum or minimum. Depending on the problem, there may be a range of coefficient parameters where the ground state is not the correct answer. Obtaining the optimal parameters to obtain the correct answer and having the ground state be the incorrect answer are contradictory, so this is not possible, but we will note this just to be safe.

From the specified value, the optimal coefficient parameters are obtained by observing the increase or decrease in the total energy and the increase or decrease in the eigenvalue difference near the ground state while adjusting the coefficient parameters. Then, a QUBO is generated near the obtained parameters to obtain consistency with the convergence probability in D-Wave.

\subsection{QUBO generation parameters and QA convergence for graph coloring problems}\label{subsec3-1}
Here, in order to see the relationship between the QUBO generation parameters of the graph coloring problem and the QA convergence, we set the coefficient parameters of the two terms in Eq. (51) of the Lucas paper to $A$ and $B$, respectively.
\begin{equation}
H = A \sum_{v} \left( 1 - \sum_{i=1}^{n} x_{v, i} \right) ^2 + B \sum_{(u v) \in E} \sum_{i=1}^{n} x_{u, i} x_{v, i}.
\label{eq5}
\end{equation}

From the equation, let's first consider the energy near the ground state. In the ground state, the nodes are painted in three different colors, one for each node. The first term only comes into effect when a color is removed or added. When a color is removed, there is no penalty due to edges, but since the number of colors becomes $0$, the value in parentheses of the first term becomes 1 and the energy increases by $A$. When a color is added, in addition to $A$, the nodes connected by edges will also become the same color, and the energy increases significantly to $B \times N/3$. As a result, when $N$ is large and the balance of $A$ and $B$ is not significantly disrupted, the energy difference near the ground state is denoted by $A$.

It was found that the gap near the ground state is determined by $A$, but in the actual QA machine , auto-gain control is used to promote the time evolution of the Schr\"{o}dinger equation as much as possible. In other words, the limit is the dynamic range, which is the ratio of the gap near the ground state to the maximum and minimum eigenvalue differences, so we consider the maximum values of the first and second terms. The magnitudes of the first and second terms are maximum when all colors are selected for each node. The first term increases linearly with increases in $A$ and $N$, but increases squared with increases in the number of colors $k$. The second term also increases linearly with $B$, but increases in proportion to the increase in the number of edges in the complete Department graph, $N/k \times N(1-1/k) \times k/2$, so it increases as the square of the number of nodes. This does not apply to problems with no limits such as the 3-color problem. Basically, as the number of edges increases, the number of colors required also increases, and the number of QUBO variables, spins, and QUBO coefficients also increase. A similar discussion can be made by finding the maximum and minimum of the generation formula according to the conditions at that time.

\begin{figure}[t]
\centering
\includegraphics[width=0.9\textwidth]{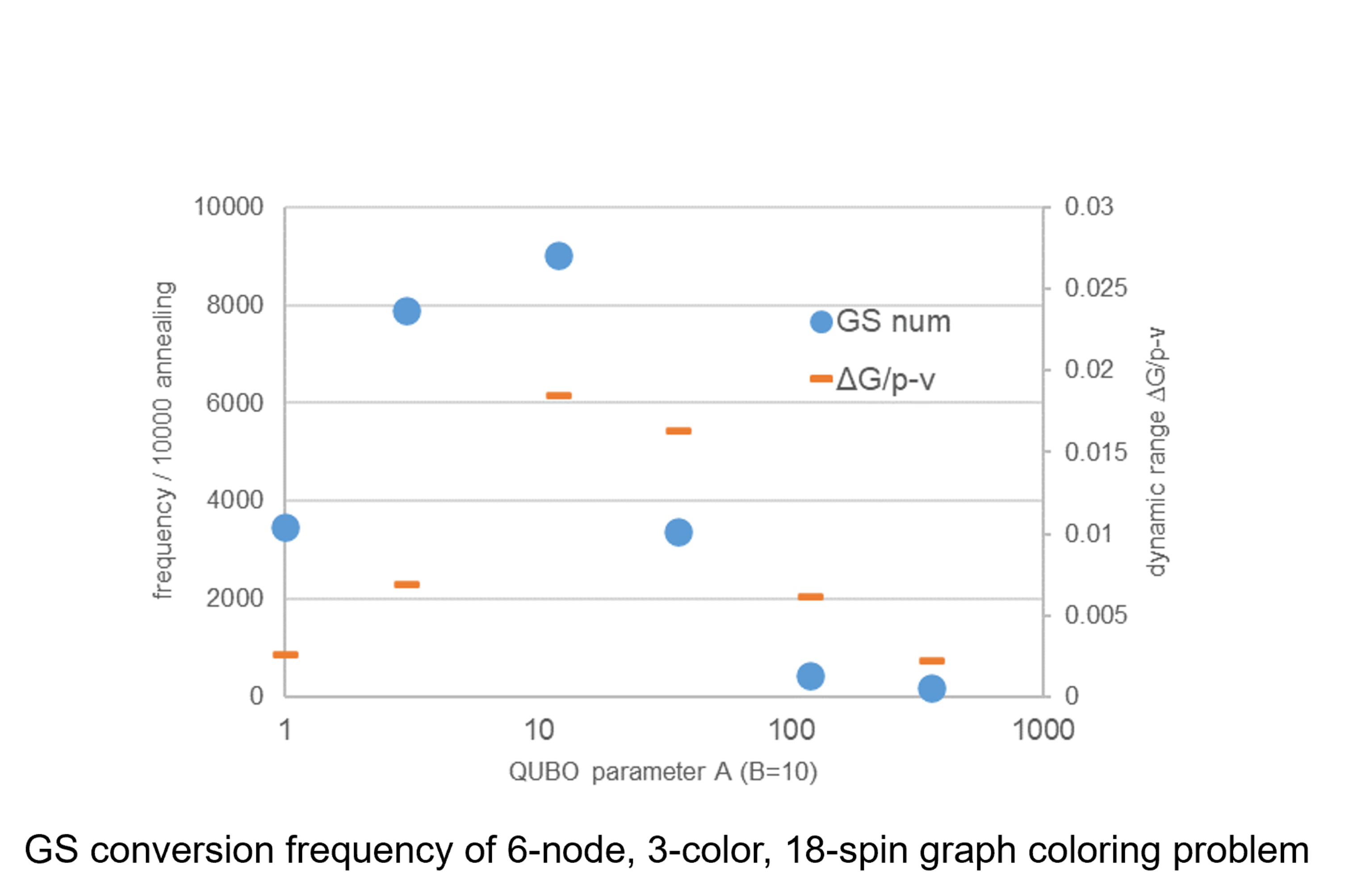}
\caption{The horizontal axis is the value of the first term coefficient parameter $A$ when generating the QUBO. $B$ is fixed at 10. The circles are the sum of the ground state convergence numbers for 10 iterations of annealing 1000 times for each embedding of the D-Wave QPU and QUBO coefficients on the graph. This is the total for 10 types of embedding. The horizontal line is the dynamic range, which is the value obtained by dividing the energy difference near the ground state obtained from equation (5) by the maximum energy value.}
\label{fig3}
\end{figure}

Here, we fixed the second term coefficient parameter of the QUBO for the 6-node 3-color problem at $B=10$, and varied the first term alphanumeric parameter from 1 to 360. Figure \ref{fig3} shows the ground state convergence frequency when quantum annealing was performed using QUBO on D-Wave with autoscaling on, and setting the ratio of the energy gap $A$ to the maximum eigenvalue difference (dynamic range). 

The horizontal axis corresponds to the change in $A$, and the vertical axis on the left corresponds to the solid circle, which is the number of times that the system converged to the ground state out of 10,000 20-usec annealings on the D-Wave QPU (1000 annealings were performed 10 times with different embeddings for each automatic embedding). The problem is a 6-node 3-color problem, which is a small-scale problem that becomes an 18-spin problem when converted to QUBO. But it can be seen that the convergence probability to the ground state deteriorates significantly when the balance of parameters is significantly changed. In addition, the figure shows that the maximum is approximately where the dynamic range is large. The asymmetry of the size of the dynamic range is thought to be due to the potential shape associated with the change in $A$. When the number of spins is small, a certain degree of convergence probability to the ground state can be obtained even if the parameters are changed, but the convergence probability decreases as the problem scale is increased.

Tables \ref{tab1} and \ref{tab2} show the results for problems of two sizes. Tables \ref{tab1} shows the number of ground state convergences per 1000 trials on D-Wave for the 42-node, 126-spin problem, and tables \ref{tab2}  shows the number of ground state convergences per 1000 trials on D-Wave for the 60-node, 180-spin problem. Looking at Table \ref{tab2} in comparison with Table \ref{tab1}, we can see that for the 42-node problem, the ground state convergence number depends only on the ratio of the two coefficients, and is almost constant for each ratio. For the larger problem size of the 180-spin problem, the effect of embedding becomes more pronounced and the changes become larger. For all of the problems used in this study, when multiple runs were performed, the standard deviation was roughly about the average number of convergences. The results are also shown for the clique vertex cover problem in the next section. With Pegasus, D-Wave's basic QPU, ground states were obtained for problems up to 63 nodes and 189 spins, and at that time, the ground state was obtained only when $A : B$ was $2:1$. Table 1 shows an example, but for problems with 30 nodes and 90 spins or more, the probability of convergence to the ground state is highest when $A : B$ is $4:1$. Using Zephyr, a new D-Wave QPU with improved performance, the convergence probability improved by about 10 times. However, the maximum embedding capacity according to the specifications became smaller, up to 42 nodes and 126 spins. Even with Zephyr, the convergence probability to the ground state was highest when $A : B$ was $2:1$.
\begin{table}[t]
\caption{Convergence number of 42-nodes GCP @ 1000 times. It can be seen that the ratio of the first and second term coefficient parameters determines the ground state convergence frequency.}\label{tab1}%
\begin{tabular}{@{}r|cccc@{}}
\toprule
$A$ \textbackslash \ $B$ & 2 & 4 & 8 & 16\\
\midrule
$2B$   & 81 & 99 & 108 & 39 \\
$4B$   & 473 & 426 & 514 & 285 \\
$8B$   & 89 & 74 & 71 & 71 \\
$16B$  & 0 & 0 & 0 & 0 \\
\botrule
\end{tabular}
\end{table}
\begin{table}[t]
\caption{Convergence number of 60-nodes GCP @ 1000 times. It can be seen that the ratio of the first and second term coefficient parameters affects the ground state convergence frequency, but there is a large variability.}\label{tab2}%
\begin{tabular}{@{}r|cccccc@{}}
\toprule
$A$ \textbackslash \ $B$ & 32 & 64 & 128 & 256 & 512 & 1024\\
\midrule
$2B$   & 118 & 0 & 0 & 7 & 0 & 0  \\
$4B$   & 93 & 8 & 166 & 0 & 73 & 0 \\
$8B$   & 0 & 0 & 0 & 103 & 0 & 149 \\
$16B$  & 0 & 0 & 0 & 0 & 0 & 0  \\
\botrule
\end{tabular}
\end{table}

\subsection{Clique vertex coverage problem}\label{subsec3-2}
In the clique vertex cover problem, unlike the graph coloring problem, the coefficients of the two terms in Eq. (\ref{eq2}) are used as is because the parameters are set independently from the beginning. The first term is a penalty term that assigns one color to one node, the first term within the second term is the number of edges when the node is specified as a clique. The second term is the number of edges between the extracted nodes, which is zero only when the extracted node forms a clique, and a positive integer when it does not. In other words, just like the graph coloring problem, the correct answer is when both terms are $0$.

Here, as with the graph coloring problem, we created a clique vertex cover problem with a known solution as follows. As mentioned before, multiple independent cliques are connected by only one edge. It is obvious that this is a trivial problem if the scale is small.

\begin{figure}[t]
\centering
\includegraphics[width=0.9\textwidth]{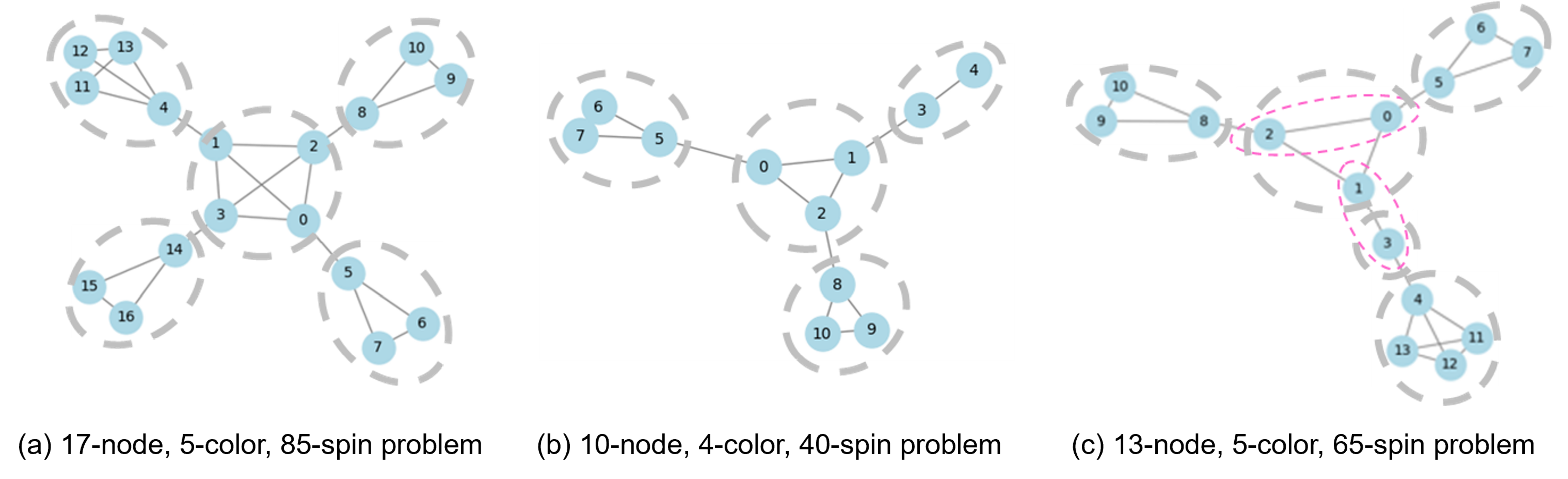}
\caption{Examples of clique vertex cover problems with known solutions. From the left, there are a 17-node, 5-color, 85-spin problem, a 10-node, 4-color, 40-spin problem, and a 13-node, 5-color, 65-spin problem.}
\label{fig4}
\end{figure}

Figure \ref{fig4} shows the examples range from the 16-node, 5-color, 80-spin problem on the left to the 10-node, 4-color, 40-spin problem in the center, and to a problem with degeneracy, 13-node, 5-color, as shown on the right. The tables below are, from top to bottom, examples of the 14-node, 4-color, 56-spin problem in Table \ref{tab3}, the 14-node, 5-color (with degeneracy) 70-spin problem in Table \ref{tab4}, and the 17-node, 5-color, 85-spin problem in Table \ref{tab5}. One embedding involved 1000 annealings, which were repeated 10 times. The graph shows the average and standard deviation of the number of times the ground state converged for each 1000 annealings. The top row shows the parameter values of $A$ and $B$, p-v represents the maximum and minimum eigenvalue difference, ave represents the average, and SD represents the standard deviation. It is observed that when the convergence frequency is low, the average and standard deviation values are almost comparable, and the optimal value is obtained with a parameter ratio of approximately $1:1$.

\begin{table}[b]
\caption{14-node, 4-color, 56-spin clique vertex coverage problem}\label{tab3}%
\begin{tabular}{@{}r|ccccccc@{}}
\toprule
$14n4c$ & $A3B1$ & $A2B1$ & $A1B1$ & $A1B2$ & $A1B3$ & $A1B5$ & $A1B10$\\
\midrule
p-v   & 658 & 532 & 406 & 686 & 966 & 1526 & 2926  \\
ave   & 3.1 & 12.2 & 31.2 & 30.2 & 21.5 & 11.5 & 1.3  \\
SD   & 2.1 & 3.3 & 5.0 & 11.3 & 7.6 & 7.6 & 1.8  \\
\botrule
\end{tabular}
\end{table}

\begin{table}[b]
\caption{14-node, 5-color (with degeneracy), 70-spin clique vertex coverage problem}\label{tab4}%
\begin{tabular}{@{}r|cccc@{}}
\toprule
$14n5c$ & $A1B1$ & $A1B2$ & $A1B3$ & $A1B5$\\
\midrule
p-v   & 584 & 944 & 1304 & 2024  \\
ave   & 10.5 & 9.9 & 5.3 & 0.8  \\
SD   & 3.1 & 8.2 & 3.1 & 1.3  \\
\botrule
\end{tabular}
\end{table}

\begin{table}[t]
\caption{17-node, 5-color, 85-spin clique vertex coverage problem}\label{tab5}%
\begin{tabular}{@{}r|cccc@{}}
\toprule
$14n5c$ & $A2B1$ & $A1B1$ & $A1B2$\\
\midrule
p-v   & 1099 & 827 & 1382 \\
ave   & 0.3 & 1.2 & 0.0 \\
SD   & 0.5 & 1.8 & 0.0 \\
\botrule
\end{tabular}
\end{table}

\subsection{Parallel machine scheduling problems}\label{subsec3-3}
In Lucas's paper, the parallel machine  problem is treated in the same section as the previous two problems, but we will see that this problem is unique and the parameter balance changes significantly:
\begin{eqnarray}
H &=& \sum_{i=1}^{N} L_{i} x_{i, 1} + A \sum_{i=1}^{N} \left( 1 - \sum_{\alpha=1}^{m} x_{i, \alpha} \right) ^2 \nonumber \\ 
&+& B \sum_{\alpha=2}^{m}  \left(\mathcal{M} - \sum_{i=1}^{N} L_{i} (x_{i, 1} - x_{i, \alpha})- \sum_{n = 0}^{\lfloor \log_2 (\mathcal{M}-1)\rfloor} 2^n z_n \right) ^2 .
\label{eq6}
\end{eqnarray}

In the formula, the first term represents the longest process time among all the machines, which is minimized under the constraints defined by the second and third terms. The variables are defined as follows: $A$ and $B$ are parameters used to adjust the weight of each term, $i$ denotes a job, $N$ is the total number of jobs, $\alpha$ represents a machine, $m$ is the total number of machines, $\mathcal{M}$  refers to the allowable process time difference between the machines with the longest process times and others, $L$ indicates the process time of each job, and $z_n$ represents an auxiliary spin used for the Log encoding of process time differences, as described in \cite{Lucas14}.
By setting the coefficient of the first term in the formula to 1, the degree of freedom for the coefficient parameters used in the second and third terms becomes 2. As the discussion so far has shown, it is important not to unnecessarily increase the dynamic range of the eigenvalues in terms of solvability, so we will first focus on the third term. This term includes the process length of each job times the maximum number of machines. In other words, there is a state in which one job is assigned to all machines at most. Therefore, it can be seen that the maximum value depends entirely on the number of jobs and process length in the problem. In the previous two sections, when we discussed the energy, \textit{i.e.}, the maximum eigenvalue, it was shown that it is possible to assign all colors to one node or paint it with multiple colors by expressing the state as spin, and that in that case, the energy is maximized. The maximum value here requires some consideration, but it is almost the same idea. The difference is that the third term is maximized when no jobs are assigned to the machine specified to maximize the total process time, and the fact that all jobs are assigned to other machines, as seen in the previous example. In this case, the first term becomes zero, and the second term, excluding the parameters that are coefficients becomes $N \times (m - 2) ^{2}$. The third term, excluding the parameters that are also coefficients, becomes $(m - 1) \times (\mathcal{M} + \sum_{i=1}^{N} L_{i}) ^{2}$. In other words, the magnitude of the third term usually dominates the overall energy. The vicinity of the ground state is occupied by one job, one machine assignment, and the process time difference between the machine with the longest process time and other machines. Since the length of the job is a natural number, the process time between machines is usually smaller than the minimum process time of the job, and since the coefficient parameter is fixed to 1 here, the gap near the ground state is fixed even if the coefficient parameters of the second and third terms are changed. Considering that the third term is usually large, it is better to have a small coefficient parameter for the third term unless there is an inconvenience that the ground state does not have the minimum makespan when generating QUBO. When the difference between the total process time and the maximum designated machine is reversed, the third term is a squared term, so if $B$ in Eq. (\ref{eq6}) is around 2, the squared term will be effective and it will be effective as a constraint term. If the number of jobs is $N$, the number of machines is m, the maximum allowable process time between machines is $M$, and $\sum_{i=1}^{N} L_{i}$ is the total process time of all jobs, then when the eigenvalue is maximized,
\begin{equation}
\text{max} E = 0 + A N (m - 2)^2 + B (m - 1) (\mathcal{M} + \sum_{i=1}^{N} L_{i})^2 .
\label{eq7}
\end{equation}

However, as will be explained later, this is the expression before $A$ becomes large and the second term becomes larger than the third term. Basically, it will be shown later that in order for the second term to become larger than the third term, the value of the coefficient parameter of the second term must be very large.

We will consider the relationship between the minimum makespan solution for the parallel machine problem and the lowest energy state in Eq. (\ref{eq7}). What makes this problem different from the previous two problems is that there exists a state that shows the minimum makespan other than the ground state. Because there is a constraint equation, the ground state cannot be obtained unless the machine with the longest process time in the first term has a process time greater than or equal to the other machines. However, if we consider this simply as the parallel machine problem before making it into QUBO, there is no need to specify the machine with the longest process time. The minimum makespan is still achieved for the parallel machine problem even if the set of jobs assigned to each machine that has reached the ground state in the above equation is arbitrarily swapped between machines.

We consider the case where the coefficient parameter $A$ of the second term is increased. Basically, near the ground state, if $A$ is a certain appropriate size, the second and third terms are $0$, and we consider the appropriate minimum value of $A$. If $A$ is smaller than the job's process time, for example, the decrease in the first term becomes larger than the penalty of not assigning a machine to the job in the second term, and the constraint condition does not hold. It must be at least larger than the maximum process time of the job. The results obtained later show that the optimal value of the coefficient parameter of the second term is quite large, so we will not discuss it further here. As discussed earlier, the coefficient parameter of the first term is fixed at 1, so even if the two coefficient parameters are changed, the energy difference near the ground state does not change in the case of parallel machine problems. In other words, the dynamic range is inversely proportional to the maximum energy value. Therefore, excluding the coefficient parameters, the second and third terms are $N \times (m - 2)^2$ and $(m - 1) \times (\mathcal{M} + \sum_{i=1}^{N} L_{i}) ^{2}$, but the maximum value of the second term alone without considering the third term, rather than considering the maximum total eigenvalue, is $N \times (m - 1)^2$ . If we fix the coefficient parameter of the third term to 2 and consider how it changes when the coefficient parameter of the second term is increased, it is thought that the optimal parameters can be determined in the same way as for the graph coloring problem and the clique vertex cover problem.

\begin{figure}[t]
\centering
\includegraphics[width=0.9\textwidth]{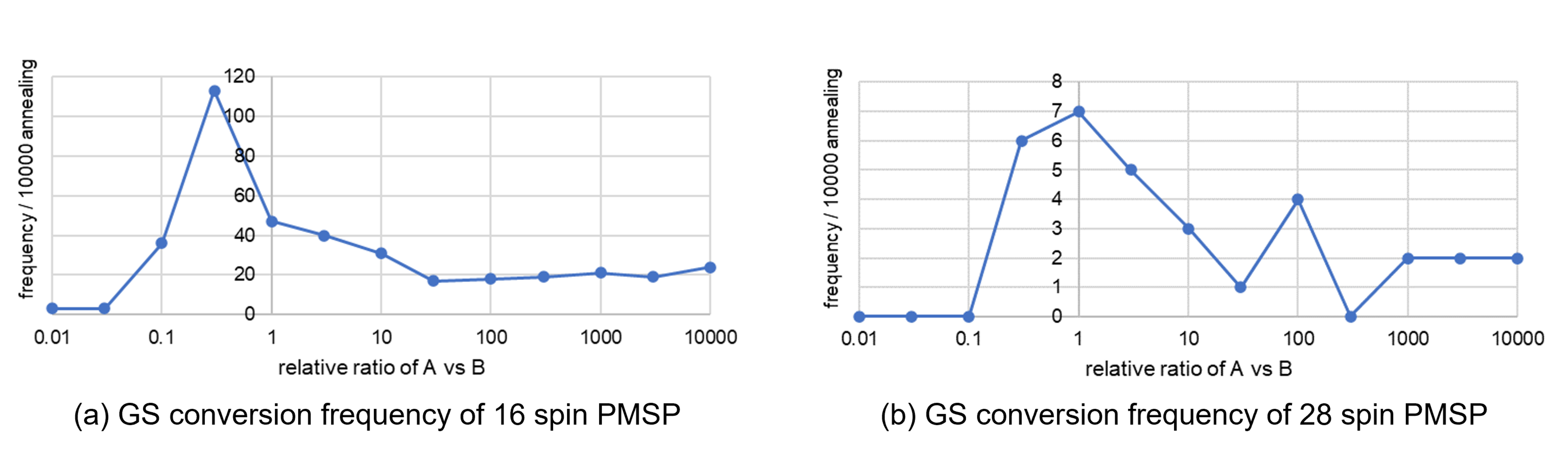}
\caption{A graph plotting the number of D-Wave ground state convergences versus the ratio of normalization coefficient parameters on the horizontal axis. A phenomenon seen in comparing Tables 1 and 2 is that when the number of spins is small and the convergence probability is reasonably high, the effect of embedding is small and the convergence number changes stably (a). As the number of spins increases and the convergence probability approaches zero, the effect of embedding becomes large, causing the convergence number to vary more (b).}
\label{fig5}
\end{figure}

Figure \ref{fig5} shows the number of times the D-Wave QPU reaches the ground state when the horizontal axis represents the ratio of the magnitude of the second and third terms when the parameter of the second term is changed.

$B$ is fixed at 2 and $A$ is set, and the horizontal axis shows the ratio of the entire second term to the entire third term. The left is a two-machine, 6-job problem with element job process times of 19, 13, 12, 21, 16, and 7. $M$ is set to a large value of 15, resulting in a 16-spin problem. The minimum makespan is 44 for both machines, and is 46 and 42 when solved using the greedy method. The right is a two-machine, 12-job problem with $M$ also set to a large value of 15, resulting in a 16-spin problem. The jobs are 73, 71, 59, 47, 41, 37, 79, 67, 61, 53, 43, and 25. There is only one correct combination for the 16-spin problem, and the degeneracy of the ground state is 2. The 28-spin problem has 10 combinations, resulting in a degeneracy of 20.

In both of these problems, the elements are two-digit natural numbers, so the total process times are large at 88 and 656, respectively, and therefore $A$, when the second term is nearly equal to the third term, is extremely large at 3540 and 75040, respectively. Looking at the graph, we can see that it hardly converges at 1/100 of that value, so in problems with weights, you need to pay close attention to the coefficient parameters.

Assigning multiple colors to one node results in redundancy due to encoding into spin, so the search space includes states that would not have been necessary if the problem had not been converted to QUBO. The advantage of quantum annealing is its fast sampling near the ground state, and the makespan can be calculated from the sampled state in a short time. In other words, if we can collect minimum makespan states for non-ground states near the ground state, and reduce the incorrect states from near the ground state, we can obtain the correct answer as a parallel machine problem by sampling the minimum makespan state near the ground state. It will be shown later that taking a large coefficient parameter for the second term makes it easier to sample realistic assignment states by widening the difference between a state in which one job is assigned to only one machine and a state in which it is not. Considering the dynamic range, it will be shown later that expanding the magnitude of the second term to the extent that it approaches the magnitude of the dominant third term is advantageous in finding a realistic solution.

Let us consider the QUBO generation formula and the size of $A$ at which the ground state is obtained. If $A$ is taken to be sufficiently large compared to the first term, and realistic job-to-machine assignment is not occurring, the eigenvalues of states other than the one-job, one-machine assignment suddenly become larger by a natural number multiple of $A$, and are far removed in energy from the ground state. In other words, only realistically assigned states exist near the ground state. It is possible that the convergence probability can be expected to be similar to or higher than that of the ground state.

\begin{figure}[t]
\centering
\includegraphics[width=0.9\textwidth]{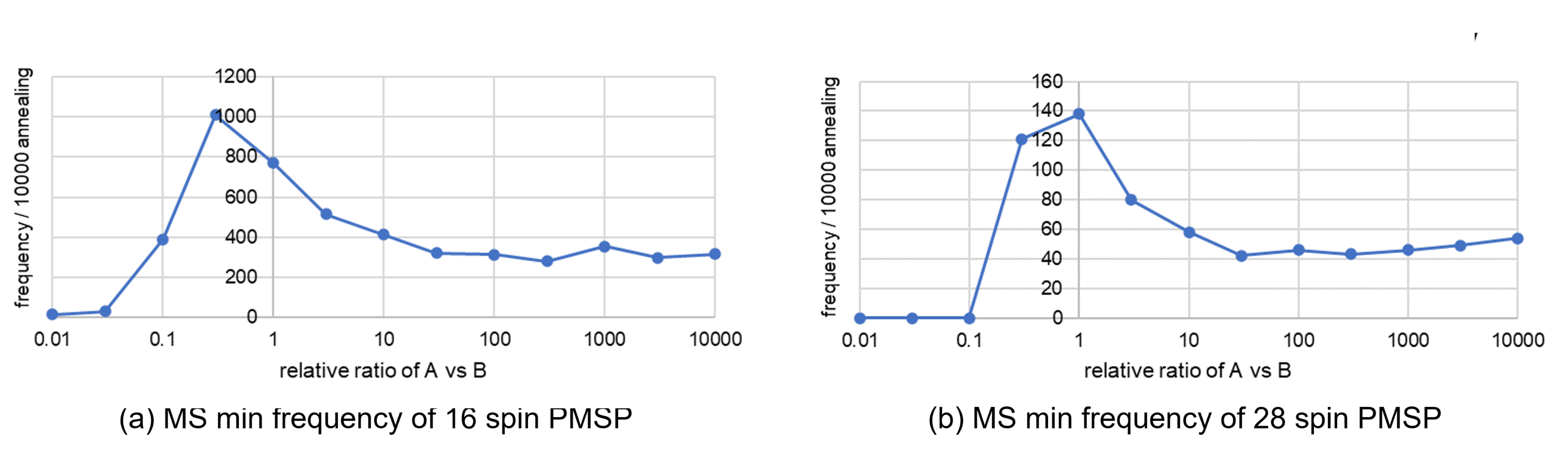}
\caption{Minimum makespan number of states. The vertical and horizontal axes are the same as in Fig. \ref{fig5}. The number of states that can be used as correct solutions is much greater than the number of ground states, with the maximum value on the vertical axis being 10 times greater in (a) and 20 times greater in (b).}
\label{fig6}
\end{figure}

As can be seen from Fig. \ref{fig6}, when the minimum makespan state is extracted, the number of convergences is about 10 times higher for the 16 spin problem, which originally had a high number of ground state convergences, but the number of convergences increases by about 20 times for the 28 spin problem, which had a low number of ground state convergences. This shows that the more difficult the problem, the more effective it is to search for states that are not ground states but that show the minimum makespan in order to find the correct answer.

In the graph coloring problem and the clique vertex cover problem, the ground state convergence probability was high when the coefficient parameter terms were well balanced. In the parallel machine problem, in order to see the effect of $A$ in the small number of spin problems, \cite{Karp72,Yarkoni22,Pinedo16,Verteletskyi20,Johnson11,Farhi00}, $M$ is set to a large value of 3, and the potential diagrams of three types of parameter balance are shown, plotting all states and eigenvalues.

\begin{figure}[b]
\centering
\includegraphics[width=0.99\textwidth]{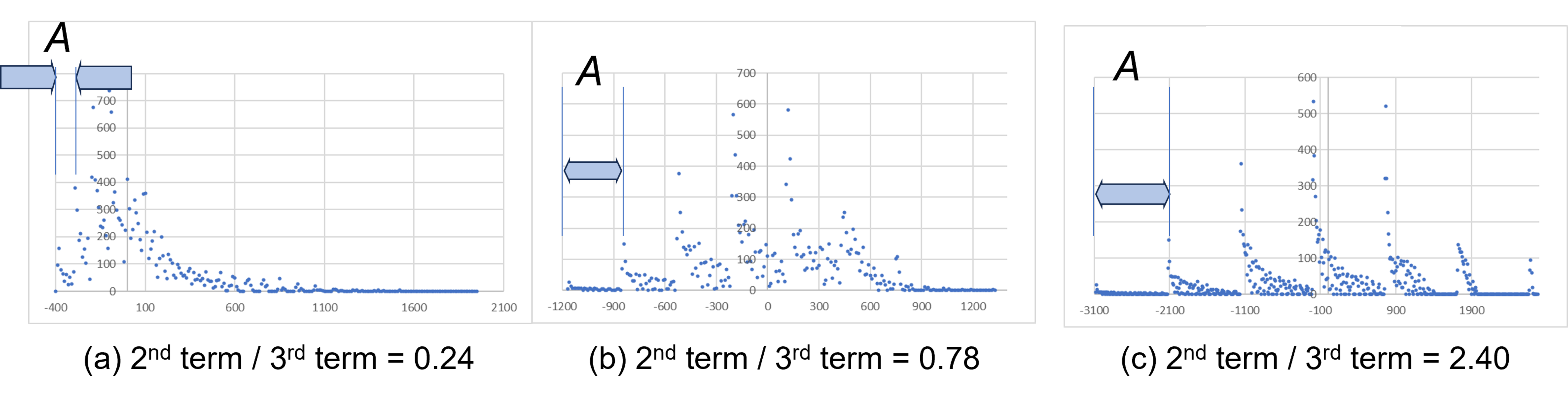}
\caption{Histogram showing eigenvalues on the horizontal axis and the number of states corresponding to the eigenvalues on the vertical axis when the coefficient parameters are changed}
\label{fig7}
\end{figure}

In Fig. \ref{fig7}, the horizontal axis is the eigenvalue, with the left end being the ground state. The vertical axis is the number of states, with a total of 16,384 states for a 14-spin problem. The states and eigenvalues are calculated after conversion to QUBO. When Eq. (\ref{eq6}) is expanded, the zeroth order of the spin, in other words, the lowest QUBO energy returned from D-Wave, does not express include constant terms in Eq. (\ref{eq6}) that are not coefficients of a spin or spins, so the ground state is negative by that amount. The magnitudes of the first and second terms in the constraint equation are in the middle, 0.78, which is close to the optimal value, with $A$ fixed at 320 and B fixed at 2. From there, the graph goes up and down by about half a digit.

The role of $A$ is the strength of the penalty for assigning one job to one machine, and in the case of the two-machine problem, the number of assigned machines is only 0, 1, or 2, so in the case of $k$ jobs other than the realistic one-machine assignment, the $A \times k$ energy rises. Looking at the histogram on the right, where A for $\text{1st} / \text{2nd} = 2.4$ is set to a large value of 960, we can see the ground state and a total of seven peaks to the right of it, spaced at intervals of $A$. This shows the effect of $A$ on the potential shape.

We will explain the effect of the potential shape, especially the maximum/minimum eigenvalue difference, on the ground state convergence number. Looking at $\text{1st} / \text{2nd} = 0.24$, the density of states near the ground state is high, with eigenvalues ranging from a minimum of -489 to a maximum of 1945, and the eigenvalue width being 2434. At $\text{1st} / \text{2nd} = 0.78$, the density of states near the ground state is much lower than at $\text{1st} / \text{2nd} = 0.24$, but the eigenvalues range from a minimum of -1161 to a maximum of 1323, and the eigenvalue width is 2484, which is almost the same as when $A$ is small. In other words, the dynamic range is almost the same, but as mentioned before, it is easier to explain if we reduce the density of states near the ground state, that is, by pushing incorrect states to higher energy, the probability of finding the ground state when converging to low energy increases. At the same time, the convergence probability to states that are not ground states but include correct states also improves.

Even with $\text{1st} / \text{2nd}= 2.4$, the density of states is low, but the eigenvalues are from a minimum of -3081 to a maximum of 2713,  and the eigenvalue difference near the ground state remains at 1, so the dynamic range also doubles. This can be understood as the reason why the increase in dynamic range is causing the convergence probability of the D-Wave QPU to decrease.

It can be seen that the condition for using this parameter effect is when the second term is much larger than the first term. In the traveling salesman problem in Section 7.2 of Lucas, the weight of the HB edge in Eq. (57) is large, that is, when there are routes with large differences in distance, it can be seen that it may be effective to take a large coefficient parameter in Eq. (56).

\section{Spin-polarity-emphasized Ising coefficients}\label{sec4}
Let's consider the mechanism for assigning one job to one machine using the Ising coefficient. Consider a system consisting of $N$ fully connected spins with an antiferromagnetic interaction coefficient. If $J_{ij} > 0$, the lowest energy is where the spin polarities are balanced, and if $N = 2$, then the lowest energy is when one is positive and one is negative. If $N = 3$ or more, then for the ± equal area, the Ising coefficient for the linear term $h_{i} > 0$ should be set so that energy rises when the spin polarity is positive. This is shown in the diagram.

\begin{figure}[t]
\centering
\includegraphics[width=0.99\textwidth]{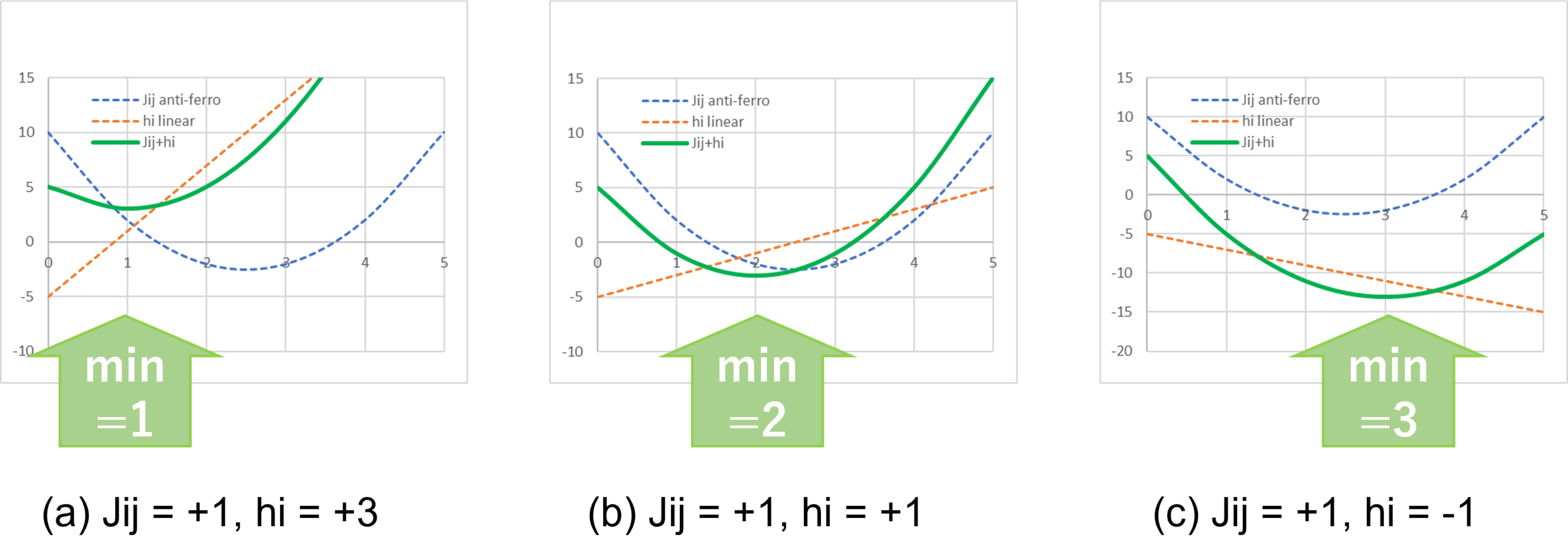}
\caption{Graph explaining how to lower the energy of the desired spin polarity. If the $N$-body interaction $J_{ij} = 1$, then by setting $h_{i} = N-2$, the energy will be lowest when only one of the $N$ bodies is positive. The above is an example of five spins, and by changing $h_{i}$, the energy can be reduced to the lowest when the positive polarity of the five spins is 1, 2, or 3.}
\label{fig8}
\end{figure}

When the spin polarity of the ground state is known, adding these is values for the Ising coefficients lowers the energy of the state of that polarity as shown in Fig. \ref{fig8}. It can be called the spin positive-negative ratio enhance Ising coefficient. However, the more biased the spin polarity is, the fewer the number of states to begin with, making it easier to find, and the more even the spin polarity is, the more exponentially the number of states increases, making it less profitable. This is particularly effective in cases such as one job, one machine assignment. This effect is similar to restricting the dimensions of a smaller area to be searched within the extra-wide dimensions expanded by QUBO and searching within that area.

\section{Discussion and conclusion}\label{sec5}
For the three types of problems in Section 6 of Lucas, we investigated the relationship between the states and eigenvalues obtained from the QUBO generation formula when changing the coefficient parameters when generating QUBO, and clarified the impact on the convergence of the QA using the D-Wave Annealer. In the graph coloring problem and the clique vertex cover problem, where the weights of the elements are uniform, the balance of the coefficient parameters does not deviate significantly from 1. For parallel machine problems, where the weights of the elements are different, the balance is greatly disrupted at the optimal value of the coefficient parameters, so caution is required, but it was shown that it is possible to consider it as a balance between the coefficient parameters, again by taking the weights into consideration. As the problem scale becomes larger and the probability of obtaining the correct ground state decreases, there was a tendency for the probability of obtaining the ground state to become sensitive to the balance of the coefficient parameters, even if the weights
\begin{equation}
T = O\left[ \exp(\alpha N^{\beta}) \right] .
\label{eq8}
\end{equation}
of the elements are uniform, so it seemed practical to search within a range of about half a digit. In addition, by being able to grasp the optimal parameters for QA convergence in advance, it is possible to reduce uncertain factors when evaluating the superiority over classical problems and make comparisons that take advantage of the high speed of QA. The probability of convergence to the ground state was calculated by taking into account the degree of degeneracy and using the formula (2) in the Lucas paper (9).

In the case of problems of 5 or greater sizes, the problem size $N$ is replaced by the number of spins:
\begin{equation}
\text{GS}_{\text{probability}} = O\left[  \exp(- \alpha N^{\beta}) \right] .
\label{eq9}
\end{equation}

\begin{figure}[b]
\centering
\includegraphics[width=0.9\textwidth]{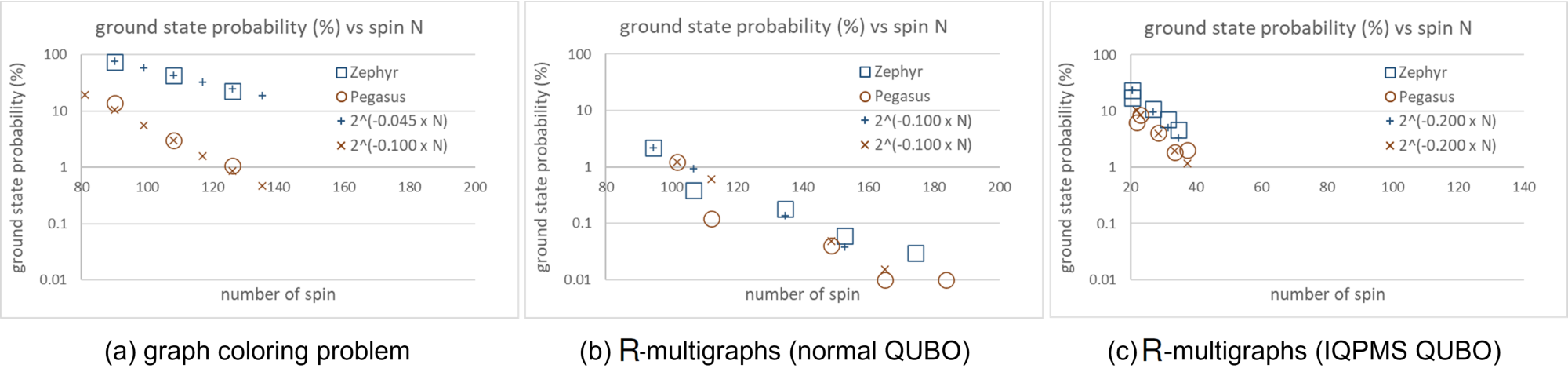}
\caption{The horizontal axis of each graph shows the number of spins when the graph problem is converted to a QUBO, and the vertical axis shows the ground state convergence probability in logarithm as a percentage when using two types of D-Wave QPUs. The convergence probability of Zephyr is shown as a hollow square, and the convergence probability of Pegasus is shown as a hollow circle. The cross and diagonal cross marks are auxiliary lines to read the slope, and the slope of each is shown in the graph. When the number in parentheses is -1, it is classical random selection, and when the slope is gentle, it can be understood that there is little decrease in the convergence probability with increasing spin number and there is a quantum acceleration element. (a) is an example of a graph coloring problem where our 3-color is the correct answer, and (b) and (c) are the annealing results of Santis et al.'s normal QUBO generation and QUBO generation with IQPMS.}
\label{fig9}
\end{figure}

When we set $\beta = 1$  and calculated $\alpha$ for each problem, we obtained $\alpha = 0.6$ for the parallel machine problem with a large dynamic range, and $\alpha = 0.1$ for the graph coloring problem, which are all neatly lined up on a straight line on the log scale when the $x$ axis indicates the number of spins. What these results show is that in graph problems where the weights are uniform, the degeneracy of the state is large, therefore the number of possible eigenvalues is much smaller than the number of states. Namely, the eigenvalue difference near the ground state is wider than in parallel machine problems with the same number of spins.  This results in a higher frequency of ground state convergence, and faster convergence times. This is consistent with the convergence theory of quantum annealing. When considering real problems, the type of real problem in which each NP problem is formulated has a large impact on quantum annealing. In addition, quantitatively understanding the differences in problems also leads to grasping the differences in convergence on the actual machine. An example is shown in Fig. \ref{fig9} and will be discussed.

It has been pointed out that the large number of spins required to express constraints when generating QUBOs places a heavy burden on embedding QUBO coefficients in a quantum annealing machine. Recently, Santis et al. showed an example in which the number of spins was compressed to about $1/4$ and the accuracy rate was improved by more than 10 times \cite{Santis24} . The convergence state at that time is shown in Fig. \ref{fig9}. (a) is an example of generating a QUBO with optimal parameters for the graph coloring problem in Section \ref{sec3}, and automatically embedding and executing it on two D-Wave processors, Pegasus \cite{Boothby19} and Zephyr \cite{Boothby21}. If we take the horizontal axis to represent spins in this way, we can see that the improvement of Zephyr is the constant offset and the decrease in the ground state convergence probability and the slowdown in the slope due to the increase in the number of spins. It is easy to qualitatively think of the former as an extension of the coherence time and the latter as an acceleration of time evolution. In the parallel machine problem, the number of spins is about 30, which is about $1/3$ of the graph coloring problem, but the effect of the latter is small. As mentioned before, the graph coloring problem has a large degree of degeneracy, a wide eigenvalue interval, and there is ample room for processor resolution, but in the parallel machine problem, the eigenvalue interval is narrow, and there is no room for processor resolution, so it is easy to understand if you think of random annealing occurring. By increasing the significant digits of the QUBO coefficients and reducing the eigenvalue difference near the ground state, the ground state convergence probability on a real machine can be reduced to almost zero even for problems with around 25 spins.

Figure 9(b) and (c) are plotted on the same graph as (a) using the same vertical and horizontal axes as (a) to compare the rate at which the ground state convergence probability decreases with increasing spin number between Pegasus and Zephyr, with the corresponding values taken from Tables 5 and 6 of the results of Santis et al. The auxiliary line is a straight line with roughly the same slope as (a). (b) shows the results using the coefficients of the normal QUBO generation process, and (c) shows the results using a QUBO generated by the combined IQPMS method, which uses both the iterative quadratic polynomial (IQP) method and the master-satellite (MS) method. To make the comparison clearer, the horizontal axis (number of spins) and vertical axis (ground state conversion probability) in all three graphs are scaled uniformly, ensuring consistent intervals. When comparing the horizontal axis of Figs. \ref{fig9}(b) and (c), it is evident that the number of spins has significantly decreased. However, the graph's slope has become steeper, suggesting that the eigenvalue intervals are narrower. Additionally, the difference between the results on Pegasus and Zephyr appears to be limited to the offset. What can be imagined from these results is that in the case of problem (c), the number of spins was reduced but the eigenvalue differences were concentrated, resulting in Zephyr's coefficient setting resolution being insufficient, and the quantum time evolution that was superior to Pegasus as seen in (a) could not be realized.
By analyzing the states and eigenvalues of various problems and comparing the execution results on actual hardware, we can evaluate the performance of quantum annealing machines on problems close to practical applications. This approach not only helps assess real-world machine performance but also promotes the practical use of quantum annealing.

We also analyzed the effects of varying the coefficient parameter $A$ in the QUBO generation constraint Eq. (\ref{eq6}) for the parallel machine problem under the "one job, one machine" constraint. After converting the QUBO coefficients into Ising coefficients, we discovered an interesting property: within any group of spins (spins used to determine which machine a job is assigned to in the parallel machine problem), there exists an independently summable Ising coefficient. This coefficient lowers the eigenvalue of a state with only one positive polarity more effectively than states with other polarities. Furthermore, we found that this property can be extended to Ising coefficients that lower the eigenvalue of a state with any specific polarity within a spin group more effectively than states with other polarities.



\backmatter





\bmhead{Acknowledgements}
The authors thank Haruki Maeda for usefull discussions for practical application.
The authors thank Juan Ivaldi for his helpful advice on understanding and representing physical phenomena.

\bibliography{sn-bibliography}

\end{document}